\documentclass{WileyMSP-template}

\usepackage{graphicx}% Include figure files
\usepackage{dcolumn}% Align table columns on decimal point
\usepackage{bm}% bold math
\usepackage[utf8]{inputenc}
\usepackage[T1]{fontenc}
\usepackage{mathptmx}
\usepackage{etoolbox}
\usepackage{chemformula} %\ch
\usepackage{amsmath}
\usepackage{cite} 
\newcommand{\JC}[1]{\textcolor{black}{#1}}
\newcommand{\JCY}[1]{\textcolor{black}{#1}}
\newcommand{\PEJ}[1]{\textcolor{black}{#1}}

\begin{document}
\pagestyle{fancy}
\rhead{\includegraphics[width=2.5cm]{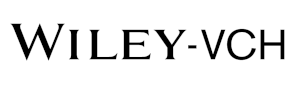}}

\title{Origin of the Apparent Electric-Field Dependence of Electrostrictive Coefficients}

\maketitle

% Author: Please give full first and last names for authors and include * after the name of all corresponding authors

\author{Jiacheng Yu}
%\author{Antoine Pautonnier}
\author{Abdelali Zaki}
\author{Killian Mache}
\author{Omar Ibder}
\author{Sandrine Coste}
\author{Maud Barr\'e}
\author{Philippe Lacorre}
\author{Pierre-Eymeric Janolin*}

% Affiliations: Please provide academic titles (Prof. or Dr.) for all authors where applicable, and include an institutional email address for all corresponding authors
\begin{affiliations}
Dr. Jiacheng Yu, Dr. Abdelali Zaki, Omar Ibder, Prof. Pierre-Eymeric Janolin*\\
Universit\'e Paris-Saclay, CNRS, CentraleSup\'elec, laboratoire SPMS, 8-10 rue Joliot Curie, 91190 Gif-sur-Yvette, France\\
Universit\'e Paris-Saclay, CNRS, CentraleSup\'elec, SPMS-Pytheas Technology SPyMS LabCom, 8-10 rue Joliot Curie, 91190 Gif-sur-Yvette, France\\
Email Address: pierre-eymeric.janolin@centralesupelec.fr

Killian Mache, Dr. Sandrine Coste, Dr. Maud Barr\'e, Dr. Philippe Lacorre \\
Institut des Mol\'ecules et Mat\'eriaux du Mans, IMMM-UMR CNRS 6283, Le Mans Universit\'e, Avenue Olivier Messiaen, F-72085 Le Mans Cedex 9, France

\end{affiliations}

% Keywords: Please provide a minimum of three and a maximum of seven keywords, separated by commas

\keywords{Electrostriction, Dielectrics, Electromechanical Materials}

% Abstract should be written in the present tense and impersonal style (i.e., avoid we), and be at most 200 words long
\begin{abstract}
Electrostrictive materials exhibit a strain that is proportional to the square of the induced polarization. In linear dielectrics where the permittivity is constant, this electromechanical strain is also proportional to the square of the electric field. However, under increasing amplitudes of the driving field, the electromechanical strain sometimes saturates;
the electrostrictive coefficients therefore appear to depend on the amplitude of the electric field used to measure them. 
Here, we present a methodology showing that this apparent field dependence is a consequence of neglecting higher-order electromechanical phenomena. When these are taken into account, not only do the electrostrictive coefficients remain constant but the signs of the high-order coefficients enable the prediction of the saturation behavior from a single measurement. We illustrate this approach on both classical and non-classical (so-called ``giant'') electrostrictors.
\end{abstract}

\section{Introduction}
Within the last decade, the interest in electrostrictive materials has gained momentum after the discovery of ``giant'' or ``non-conventional'' electrostrictors \cite{korobko2012giant,yavo2016large,makagon2021non,yu2022defining,varenik2023lead}, exhibiting electromechanical coupling constants at least ten times larger than expected from the empirical proxy\cite{newnham1997electrostriction}. 
\JC{Recently, giant effective piezoelectric constants (up to 200\,000\,pm/V at 0.1\,Hz) have been reported in thin films of Gd-doped ceria\cite{park2022induced}.  Electrostriction is of interest for electromechanical applications both as actuators and sensors. As it does not rely on the existence of ferroelectric domains, it is expected to exhibit lower losses and superior fatigue performances. A better understanding and quantification of the fundamental mechanism of the electrostrictive response will enable improvements to strengthen its potential for applications.}
Electrostriction belongs to the class of nonlinear electromechanical couplings that develop under intense electric fields. 
Such fields often correspond to moderate voltages applied to thin films or to larger samples under high voltages.
%Nonlinear phenomena of electromechanical coupling can be observed as a consequence of reduced sizes in a material system by applying a strong electric field, especially thin films. 
Besides, nonlinear electromechanical effects may also occur in materials with defects, dipole clusters, domains, etc., under moderate fields, including in ferroelectrics\cite{riemer2021dielectric,grigoriev2008nonlinear,damjanovic1998ferroelectric}. 
%For example, electrostriction \JC{(second harmonic electromechanical coupling)} occurs as a result of the opposite displacements of positive and negative ions in the crystal lattice.
%Consequently, the 
By definition, the electrostrictive strain ($x_{ij}$) is proportional to the square of the polarization ($P_kP_l$) through the electrostrictive fourth-rank tensor $Q_{ijkl}$: %and occurs at twice the
%exhibit a quadratic relation and 
%frequency of the polarization:
\begin{equation}\label{eq:defQ}
    x_{ij} = Q_{ijkl}\,P_kP_l
\end{equation}
We define ``linear dielectrics'' as insulating materials where the electric susceptibility ($\chi_e$) relating the polarization $P$ to the electric field $E$ through $P=\epsilon_0 \chi_e\,E$ is constant ($\partial \chi_e/\partial E=0$). 
In these linear dielectrics, a similar relation (Eq.\eqref{eq:defM}) relates the strain to the square of the electric field ($E_kE_l$). The corresponding electrostrictive tensor is then $M_{ijkl}$:
\begin{equation}\label{eq:defM}
    x_{ij} = M_{ijkl}\,E_kE_l
\end{equation}

However, most electrostrictive materials appear to exhibit field-dependent values of their electrostrictive tensor components rather than constant values\cite{cross1980large,varenik2020dopant,kabir2019effect,park2022induced} as expected from the definition. 
The measured electrostrictive strain is then no longer proportional to the square of the electric field or of the polarization. Most often, the electrostrictive strain saturates with increasing electric fields (\textit{i.e.}, the apparent $M$ value decreases) even though, in some of these cases, the strain nevertheless remains proportional to the square of the polarization\cite{cross1980large} (\textit{i.e.}, the $Q$ values remain constant).

Such a discrepancy between the behavior under the electric field and polarization can be rationalized as follows. 
The development of a polarization inside the dielectric material generates the electrostrictive strain.
If the polarization saturates under large electric fields, then the electric susceptibility is not linear constant anymore. 
As a consequence, the increase in polarization is not anymore proportional to the field increase and the corresponding strain, while remaining proportional to the saturating polarization squared, is not proportional anymore to the increasing field squared beyond the saturation threshold.

The more intriguing case is where the electrostrictive strain saturates as a function of both the electric field squared and the polarization squared. 
Such a case is fundamentally different from the previous one as it also occurs in linear dielectrics\cite{uchino2022electrostrictive}, which are, by definition, devoid of polarization saturation.

Discarding an electric-field dependence on the elastic properties of the materials, the question of the origin of such saturation remains and is the topic of this article. We will show that such a saturation is not related to a field dependence of the electrostrictive coefficients but rather to the influence of higher-order electromechanical phenomena.

%\JC{We should make it clear that there are two different things. The saturation of polarization induces a field-dependent electrostrictive coefficient $M$, but field-dependent of $M$ (and $Q$) may also occur in purely linear dielectrics as well\cite{uchino2022electrostrictive}, i.e., without saturation of the polarization. The is more intriguing and the topic of this article.}
  
%\JC{An} interpretation of the saturation of electrostrictive response in ``giant''  electrostrictors is presented, which enables us to draw some conclusions on the dynamics of oxygen vacancy defects\cite{zhang2023engineering}. \PE{To be reviewed at the end.}

\section{Nonlinear characterization}

Hereafter, we present a method to reliably calculate the value of electrostrictive tensor components (called ``coefficients'') from time-dependent induced strain measurements. This method enables the fitting of electrostrictive tensor coefficients that are independent of the electric field, as is expected from the definition of electrostriction. In addition, the evolution of the electromechanical strain with the electric field is explained.

\subsection{Theoretical analysis}
Similar to the definition of nonlinear susceptibilities, nonlinear electromechanical coupling between strain ($x$) and the electric field ($E$) or the polarization ($P$) can be expanded using a Taylor expansion\cite{newnham1997electrostriction}: 
%\begin{eqnarray}
\begin{equation}
    x=dE+ME^2+M^{(3)}E^3+M^{(4)}E^4+\ldots\label{eq:def1}
\end{equation}
%\end{eqnarray}
\begin{equation}
    x=gP+QP^2+Q^{(3)}P^3+Q^{(4)}P^4+\ldots\label{eq:def2}
\end{equation}
where $d$ and $g$ are piezoelectric tensors, $M$ and $Q$ electrostrictive tensors, the quadratic order of electromechanical interactions. 
The quantities $M^{(n)}$ and $Q^{(n)}$ are defined as the $n$-th ($n\geq 3$) order nonlinear electromechanical tensors.
Due to the vector nature of the electric field and polarization, $M$ and $Q$ are fourth-rank tensors, $M^{(3)}$ and $Q^{(3)}$ are fifth-rank tensors, and so on. 

Centrosymmetric dielectrics exhibit electrostriction but no piezoelectricity ($d$ and $g$ are zero). However, noncentrosymmetric dielectrics can exhibit both electrostriction and piezoelectricity, the former having been proposed as an alternative approach to improve the latter\cite{li2014electrostrictive}.
Electrostriction is, therefore, a second order electromechanical coupling instead of a secondary effect of piezoelectricity.
The contributions to the strain of the higher-order terms are not necessarily small. They depend on the magnitude of the fields and on the materials.
For example, Park \textit{et al.} reported that Gd-doped ceria thin films possess higher-order harmonics\cite{park2022induced}, which is expressed in their case through a field-dependent $M$ coefficient. We shall show that actually, such field-dependence is entirely due to the contribution of higher-order terms and the $M$ coefficient is constant.

A common electromechanical experiment is to measure the strain of a sample driven by a mono-frequency electric field:
\begin{equation}
   E(t) = E_0\sin(\omega t)\label{eq:sin} 
\end{equation}
where $E_0$ the magnitude of the driving field, $t$ the time, and $\omega$ the angular frequency. 
Combining Eq.\eqref{eq:def1} and Eq.\eqref{eq:sin}, the expression of time-dependent strain induced by a sinusoidal electric field is:
%\begin{eqnarray}\label{eq:x0(t)}
%x(t) =&& dE_0\sin(\omega t)+ ME_0^2\sin^2(\omega t)\nonumber\\
%&&+ M^{(3)}E_0^3\sin^3(\omega t)+ M^{(4)}E_0^4\sin^4(\omega t)+\ldots\label{eq:x(t)} 
%\end{eqnarray}
\begin{equation}%\label{eq:x0(t)}
    \begin{split}
        x(t) = & dE_0\sin(\omega t)+ ME_0^2\sin^2(\omega t)\\
&+ M^{(3)}E_0^3\sin^3(\omega t)+ M^{(4)}E_0^4\sin^4(\omega t)+\ldots\label{eq:x(t)} 
    \end{split}
\end{equation}
In order to fit this time-dependent strain and extract the electrostrictive coefficients, a model selection procedure could be used to determine the maximum power to consider, from a trade-off between model complexity and performance. We show below that this procedure can be advantageously replaced by using an approach based on the Fourier transform that provides additional information on the electromechanical response.

% The discrete Fourier transform of the time-dependent strain is: 
% \begin{equation}\label{eq:FFT}
%     F_k = \sum_{n=0}^{N}x_n\exp{\left(-i\,2\pi\frac{k}{N}\,n\right)}\quad k=0,\ldots,N-1
% \end{equation}
% where $N$ is the total number of measurements points and $x_n$ is the $n$-th recorded point, i.e. at a time $t_n=n.\delta t$. The exponential argument can be re-written as multiples of the excitation angular frequency $\omega$: $-i\,k\,\omega t_n$ \textit{i.e.} as harmonics. 

The discrete Fourier series \PEJ{($\{F(k\omega\}$)} of the time-dependent strain is \PEJ{defined as}:
\begin{equation}\label{eq:FFT}
    F(k\omega)=\sum_{m=0}^{N-1} x(t_m) \exp{(-i\cdot k\omega \cdot t_m)}\quad k=0,\ldots,N-1
\end{equation}
\PEJ{where $\omega=2\pi/T$ is the excitation angular frequency and $T$ its period, $N.\delta t=T$ the total acquisition time, $N$ the number of recorded points, $\delta t$ the sampling interval ($\delta t=t_{m+1}-t_m$), and $x(t_m)=x(m\delta t)$ the strain recorded at the time $t_m$}. 

%\PE{The amplitude spectrum is the set of the norm of the Fourier weights: $\left\{|F(k\omega)|\right\}_N$. }

The \PEJ{corresponding} inverse discrete Fourier series provides the expression of the strain as a function of the \PEJ{$k$-th} harmonics \PEJ{(\emph{i.e.} at $k\omega$)} of the excitation fields:
\begin{equation}\label{eq:iFFT}
     x(t) = \frac{1}{N} \sum_{k=0}^{N-1} F(k\omega) \exp{(i\cdot k\omega\cdot t)} %\quad m=0,\ldots,N-1
\end{equation}

% The inverse discrete Fourier transform therefore provides the expression of the strain as a function of the excitation field harmonics :
% \begin{equation}\label{eq:iFFT}
%     x_m = \frac{1}{N} \sum_{n=0}^{N-1} F_n \exp{\left(i\,2\pi\,\frac{m}{N}\,n\right)} \quad m=0,\ldots,N-1
% \end{equation}

%% older version
% taking real part of Eq.\eqref{eq:FFT} leads to:
% \begin{equation}
%     \mathfrak{R}\{F_k\} = \sum_{n=0}^{(N-1)/2} x_n \cos{\left(\,2\pi\frac{k}{N}\,n\right)}=\sum_{n=0}^{(N-1)/2} x_n \cos{(\omega_k\,n)}
% \end{equation}
% and Eq.\eqref{eq:iFFT}:
% \begin{equation}
%     x_m = \frac{1}{N} \sum_{n=0}^{(N-1)/2} F_n \cos{\left(\omega_m\,n\right)} \quad m=0,\ldots,(N-1)/2
% \end{equation}
% as $\omega_m = \omega\,t_m = \omega\,m\delta t$ and $x_m \equiv x(t_m)$, this can be re-written as
% \begin{equation}\label{eq:iFFT2}
%     x(t_m) = \frac{1}{N} \sum_{n=0}^{(N-1)/2} F_n \cos{\left(n\,\omega\,t_m\right)} \quad m=0,\ldots,(N-1)/2
% \end{equation}

The values of the $M$ are sometimes extracted from such a procedure, based on the Fourier transform of the strain, by only considering the component at 2$\omega$ \PEJ{(\emph{i.e.} for $k$=2 only in Eq.\eqref{eq:iFFT})}. This can be achieved through the use of a lock-in amplifier or numerically. 
We shall show below that the as-obtained $M$ values (i.e., considering that the strain at $2\omega$ is only related to the $M$ coefficient) can deviate significantly from the actual value and exhibit an artificial field dependence, even in purely linear dielectrics. Indeed, the electrostrictive coefficient $M$ is not the only one contributing to the strain at $2\omega$. 

To show this, we use the power-reduction formulae% (Eq.\eqref{eq:power-reduction}), $x(t)$ can be written as a function of multiples of the \JC{angular frequency of the exciting field (Eq.\eqref{eq:x1(t)}):} 
\begin{eqnarray}
\sin^2(\omega t) = && \frac{1-\cos(2\omega t)}{2} \nonumber\\
\sin^3(\omega t) = && \frac{3\sin(\omega t)-\sin(3\omega t)}{4} \nonumber\\
\sin^4(\omega t) = && \frac{3-4\cos(2\omega t)+\cos(4\omega t)}{8}\label{eq:power-reduction}
\end{eqnarray}
to express Eq.\eqref{eq:x(t)} as a function of multiples of the excitation angular frequency \textit{i.e.}, in terms of harmonics:
\begin{eqnarray}\label{eq:x1(t)} 
x(t) =&& \left[\frac{1}{2}ME_0^2+\frac{3}{8}M^{(4)}E_0^4+\ldots\right]+ \nonumber\\
&&\left[dE_0+\frac{3}{4}M^{(3)}E_0^3+\ldots\right]\sin(\omega t)+\nonumber\\
&&\left[-\frac{1}{2}ME_0^2+\frac{1}{2}M^{(4)}E_0^4+\ldots\right]\cos(2\omega t)+\nonumber\\
&&\left[-\frac{1}{4}M^{(3)}E_0^3+\ldots\right]\sin(3\omega t)+\nonumber\\
&&\left[\frac{1}{8}M^{(4)}E_0^4+\ldots\right]\cos(4\omega t)+\ldots
\end{eqnarray}
where the $\ldots$ stands for higher-order terms.

The %pre-factor times $\cos(2\omega t)$ 
\PEJ{third line of Eq.\eqref{eq:x1(t)}} corresponds to the strain occurring at $2\omega$\PEJ{, that is the second harmonic of the strain}. It is a weighted sum of all electromechanical couplings of even orders (\PEJ{$M\equiv M^{(2)}$, $M^{(4)}$, \ldots}) and does not only consist of electrostriction \PEJ{(\emph{i.e.} $M$ is not the only term in the prefactor)}. It is therefore only appropriate to assimilate this pre-factor to pure electrostriction if the higher-order terms \PEJ{($M^{(4)}$, \ldots)} are negligible.

To elaborate on the fact that the \PEJ{pre-factor} %Fourier weight 
corresponding to the second harmonic is not only due to electrostriction, Eq.\eqref{eq:x1(t)} can be re-written as a sum of cosines (note the sign change of the terms in some pre-factors):
\begin{eqnarray}
x(t) =&& \left[\frac{1}{2}ME_0^2+\frac{3}{8}M^{(4)}E_0^4+\ldots\right]+ \nonumber\\
&&\left[dE_0+\frac{3}{4}M^{(3)}E_0^3+\ldots\right]\cos\left(\omega t-\frac{\pi}{2}\right)+\nonumber\\
&&\left[\frac{1}{2}ME_0^2-\frac{1}{2}M^{(4)}E_0^4-\ldots\right]\cos\left(2\omega t-\pi\right)+\nonumber\\
&&\left[\frac{1}{4}M^{(3)}E_0^3-\ldots\right]\cos\left(3\omega t-\frac{3\pi}{2}\right)+\nonumber\\
&&\left[\frac{1}{8}M^{(4)}E_0^4+\ldots\right]\cos\left(4\omega t-2\pi\right)+\ldots\label{eq:x2(t)} 
\end{eqnarray}

The formulation of Eq.\eqref{eq:x2(t)} is close but not identical to the one of a Fourier series. 
%The goal is now to relate the Fourier weights (the $F(k\omega)$ in Eq.\eqref{eq:iFFT}) to the pre-factors between brackets in Eq.\eqref{eq:x1(t)}. To do so, we adopt the Fourier series formalism of Eq.\eqref{eq:iFFT}:
\PEJ{To illustrate this, the} Fourier series corresponding to Eq.\eqref{eq:iFFT} %is:
%\begin{equation}
%    x(t) = \frac{1}{N}\sum_{k=-\left\lfloor N/2\right\rfloor}^{\left\lfloor N/2\right\rfloor} F(k\omega)\exp{(i\cdot k\omega\cdot t)}
%\end{equation}
%where $\left\lfloor N/2\right\rfloor$ stands for the floor value of $N/2$ to encompass both odd and even values of $N$. 
%
%This exponential form of the Fourier series 
can also be written as an amplitude-phase form (\PEJ{as $x(t)$ is real, its transform is hermitian):}
\begin{equation}\label{eq:FFS}
    x(t) = \frac{F_0}{N} + \frac{2}{N}\sum_{k=1}^{\left\lfloor N/2\right\rfloor} |F(k\omega)|\,\cos{(k\omega\cdot t - \varphi_k)}
\end{equation}
where $\left\lfloor N/2\right\rfloor$ stands for the floor value of $N/2$ to encompass both odd and even values of $N$, 
%where
$\varphi_k$ is the phase of $F(k\omega)$ (i.e., $F(k\omega)=|F(k\omega)|e^{i\varphi_k}$) and the first term corresponds to the average value of $x(t)$ over the recorded period.

To further underline the difference between Eq.\eqref{eq:FFS} and Eq.\eqref{eq:x2(t)}, we explicitly write Eq.\eqref{eq:FFS} up to the fourth harmonic:
\begin{eqnarray}\label{eq:exp_FCoeff}
    x(t) &=& \frac{F_0}{N} + \nonumber\\
    &&\left[\frac{2|F(\omega)|}{N}\right]\cos{(\omega\cdot t - \varphi_1)} + \nonumber\\
    &&\left[\frac{2|F(2\omega)|}{N}\right]\cos{(2\omega\cdot t - \varphi_2)} + \nonumber\\
    &&\left[\frac{2|F(3\omega)|}{N}\right]\cos{(3\omega\cdot t - \varphi_3)} + \nonumber\\
    &&\left[\frac{2|F(4\omega)|}{N}\right]\cos{(4\omega\cdot t - \varphi_4)} + \cdots 
\end{eqnarray}

Therefore the presence of the phase of the Fourier weights (the $\varphi_n$'s \PEJ{in Eq.\eqref{eq:exp_FCoeff}}) instead of successive multiples of $\pi/2$ \PEJ{(in Eq.\eqref{eq:x2(t)})} prevents the identification of the normalized amplitude spectrum of the Fourier series (the $\left\{\frac{2|F(k\omega)|}{N}\right\}$ \PEJ{in Eq.\eqref{eq:exp_FCoeff}}) with the pre-factors \PEJ{(the terms between square brackets)} of Eq.\eqref{eq:x2(t)}.

\subsection{Methodology to calculate electromechanical coefficients}

\JC{Even though it is not possible to directly calculate the electrostrictive coefficients (in Eq.\ref{eq:x2(t)}) from the Fourier spectrum of the strain (Eq.\ref{eq:exp_FCoeff}),} a Fourier analysis is nevertheless useful to obtain the electromechanical coefficients, including the $M$ electrostrictive coefficient. We present below a reliable method to extract the electromechanical coefficients based on the measurements of $E(t)$ and $x(t)$ that (1) reproduces the observed saturation of the electrostrictive strain even in linear dielectrics and that (2) provides values of the electromechanical coefficients that are field-independent. 

The amplitude spectrum $\left\{\frac{2|F(k\omega)|}{N}\right\}$ of the Fourier transform of $x(t)$ as defined by Eq.\eqref{eq:FFS} is computed first. It serves as a basis for the choice of the harmonics to consider. The highest harmonic ($h_m\omega$) to consider should be set to prevent under- or over-fitting the data. 

It is indeed necessary to only keep harmonics with weights exceeding the uncertainties of the fast Fourier transform.
The substantial aliasing and spectral leakage errors that discrete Fourier transform may introduce are very sensitive to measurement noise contaminations\cite{ma1997accurate,dishan1995phase}.  
To address this issue, a method based on the Volterra series was reported for the accurate measurement of higher-order frequency response functions to identify structural nonlinearities\cite{lin2018new}.
Considering the calculation complexity, measuring entire multi-dimensional higher-order frequency response functions may not be necessary. 
For practical purposes, we shall provide a simpler description to calculate electromechanical coupling coefficients in the presence of higher-order terms: \JC{a histogram of the Fourier weights with a hundred bin is used to determine the relevant harmonics (see Suppl. Information). In our measurements, the noise level of the FFT corresponds to 1 or 2\% of the normalized Fourier weights. All harmonics corresponding to a Fourier weight above that threshold are selected.}

Once the harmonics are selected, a fit of $x(t)$ is carried out\footnote{Eq.\ref{eq:fit} gather the electromechanical coupling constants and the powers of the field in order to have coefficients of similar amplitude, to ease the fitting procedure. The $x(t)$ curve is fitted rather than the $x(E)$ curve for this reason as well, in addition to the fact that the losses complicate the fit.} with a function $f(t)$ defined as:
\begin{equation}
    f(t) = \sum_{\{h_n\}} f_n\sin^n(\omega t+\phi_n)
    \label{eq:fit}
\end{equation}
The result of the fit is the set of $\{f_n\}$ that are the $n$-th electromechanical coupling coefficients times the corresponding electric field power: $f_1=dE_0$, $f_2=ME_0^2$, and $f_{n}=M^{(n)}E_0^n$ $\forall n\geq 3$. Note that these are not the pre-factors of Eq.\eqref{eq:x1(t)} but of Eq.\eqref{eq:x(t)}. %\eqref{eq:x0(t)}. 
They therefore provide directly (once divided by the $n$-th power of the electric field) the $n$-th electomechanical coefficient.

\section{Experimental examples}

We illustrate this methodology on classical and non-classical electrostrictors to demonstrate its ability to both prove field-independent values of the electrostrictive coefficients and explain the saturation of the electromechanical strain \JCY{shown in Figure \ref{fig:Fig1}}. All measurements are carried out at 10\,Hz and 20$^\circ$C on ceramic samples of similar dimensions. 
\JCY{The measured longitudinal displacements induced by the electric field along the same direction} are of several hundred nanometers, far greater than the minimal measurable value (4\,nm) and the measured strains have been averaged over 10 cycles.

%The electrostrictive strain saturation of 0.9\ch{Pb(Mg_{1/3}Nb_{2/3})O3}-0.1\ch{PbTiO3} (PMN-10PT) ceramics and substituted \ch{La2Mo2O9} ceramics are presented in Figure \ref{fig:Fig1}.
%From the definition of electrostriction, the strain \JC{is expected to exhibit} linear dependence of the square of the external field or the polarization. 

%Hence, the generally accepted mechanism is that the polarization in ferroelectrics gives rise to the induced strain under the electric field, and the piezoelectric coefficients of these ferroelectrics originate from electrostrictive effect and spontaneous polarization. 
 
\subsection{Harmonic analysis of classical electrostrictors}

\begin{figure*}[hbtp]
    \centering
    \includegraphics[width=0.95\textwidth]{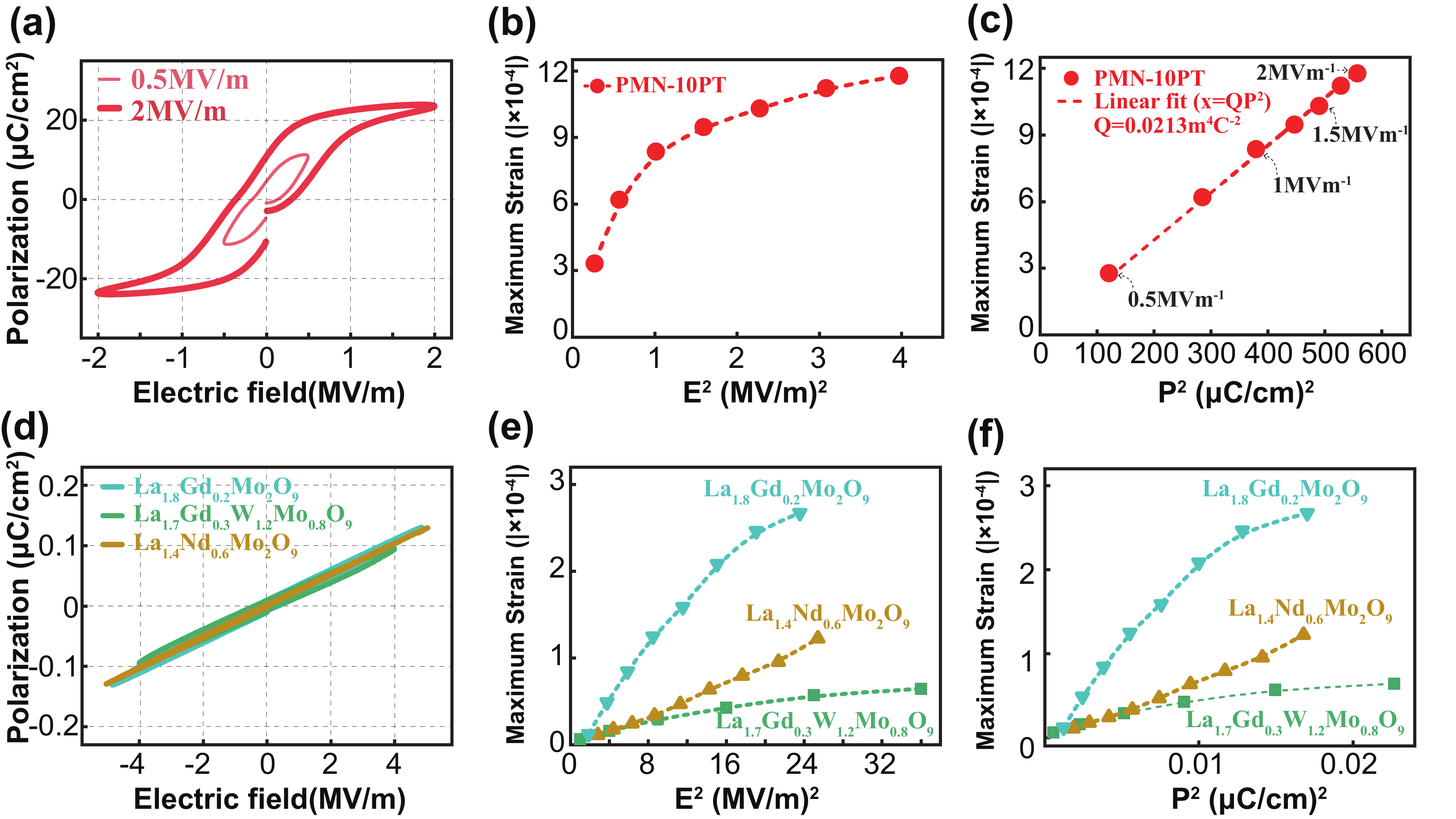}
    \caption{Polarization as a function of the electric field ($P$-$E$ loop) of 0.9\ch{Pb(Mg_{1/3}Nb_{2/3})O3}-0.1\ch{PbTiO3} (PMN-10PT) ceramics (a) and substituted \ch{La2Mo2O9} ceramics (d). Electrostrictive responses of PMN-10PT as a function of the square of electric field amplitude (b) and of the square of polarization (c). All the measurements are at 10\,Hz and 20$^\circ$C. The $x$-$E^2$ curve of PMN-10PT (b) saturates while the $x$-$P^2$ curve (c) remains linear. In contrast, substituted \ch{La2Mo2O9} are all linear dielectrics (d). As a consequence, the $x$-$E^2$ curve of substituted \ch{La2Mo2O9} (e) and $x$-$P^2$ (f) curve exhibit a similar tendency: Gd- and (Gd,W)-substituted \ch{La2Mo2O9} saturate at the large fields whereas Nd-substituted \ch{La2Mo2O9} shows the opposite behavior.}
    \label{fig:Fig1}
\end{figure*}

We call ``classical electrostrictors'' materials exhibiting electrostrictive properties described by the empirical proxy\cite{newnham1997electrostriction}.
Some of the most common ones are based on the relaxor ferroelectric lead magnesium niobate (PMN), with a maximum strain around $0.2\%$\cite{park1997ultrahigh,liu1999electric,bokov2006recent,donnelly2003dielectric} for ceramics. 
Ceramic samples of 0.9\ch{Pb(Mg_{1/3}Nb_{2/3})O3}-0.1\ch{PbTiO3} (PMN-10PT) have therefore been investigated. 
The electric-field-induced evolution of polar nano-regions (PNRs) in relaxor ferroelectrics plays an important role in their electromechanical properties\cite{burns1983crystalline,shirane2005dynamics,shi2022quantitative}.
In this PMN-10PT, the saturation of the polarization (Figure \ref{fig:Fig1}(a)) causes the saturation of the electrostrictive strain under the electric field. This is illustrated in Figure \ref{fig:Fig1}(b) where the maximum strain saturates when plotted against the electric field squared, whereas it remains linear when plotted against the polarization squared (Figure \ref{fig:Fig1}(c)).  Such behavior leads to an apparent non-constant, field-dependent $M$ coefficient, whereas the $Q$ coefficient remains constant as expected.

The curve $x$-$E^2$ (Figure \ref{fig:Fig1}(b)) for PMN-10PT ceramics may be thought of as consisting of two parts: a linear part below about 1\,MV/m and a non-linear part under higher electric field. 
Li \textit{et al.} explained similar results in PMN crystals by separating the response into such low- and high-field regions\cite{li2014effect}.
In the high-field region, the saturation of the electromechanical strain was explained by the fact that all the PNRs have been switched to a stable state, causing the saturation of both polarization and electrostriction. 
This explanation is indeed consistent with Figure \ref{fig:Fig1}(c), showing that the curve $x$-$P^2$ is approximately linear rather than saturated. 
In relaxor ferroelectrics as well as in nonlinear dielectrics, the saturation of the polarization under the electric field is indeed usually the primary reason for the saturation of the electromechanical strain.
This is the reason why electrostriction in ferroelectrics is preferably assessed by $Q$ coefficients. 
Sundar \textit{et al.} demonstrated in addition that the $Q$ coefficient is fairly independent of temperatures in PMN ceramics\cite{sundar1992electrostriction}. 
We shall however show below that this explanation omits the contribution of higher-order couplings and that such separation in low- and high-field parts is actually not necessary.

%Before explaining why the strain saturates under electric field, it is important to note that in non-linear dielectrics such as ferroelectrics, the saturation of the polarization under electric field is usually the primary reason for the saturation of the electromechanical strain. This saturation does not however entail a field-dependence of the electrostrictive $M$ coefficient as we show below.

\begin{figure*}[htbp]
    \centering
    \includegraphics[width=0.95\textwidth]{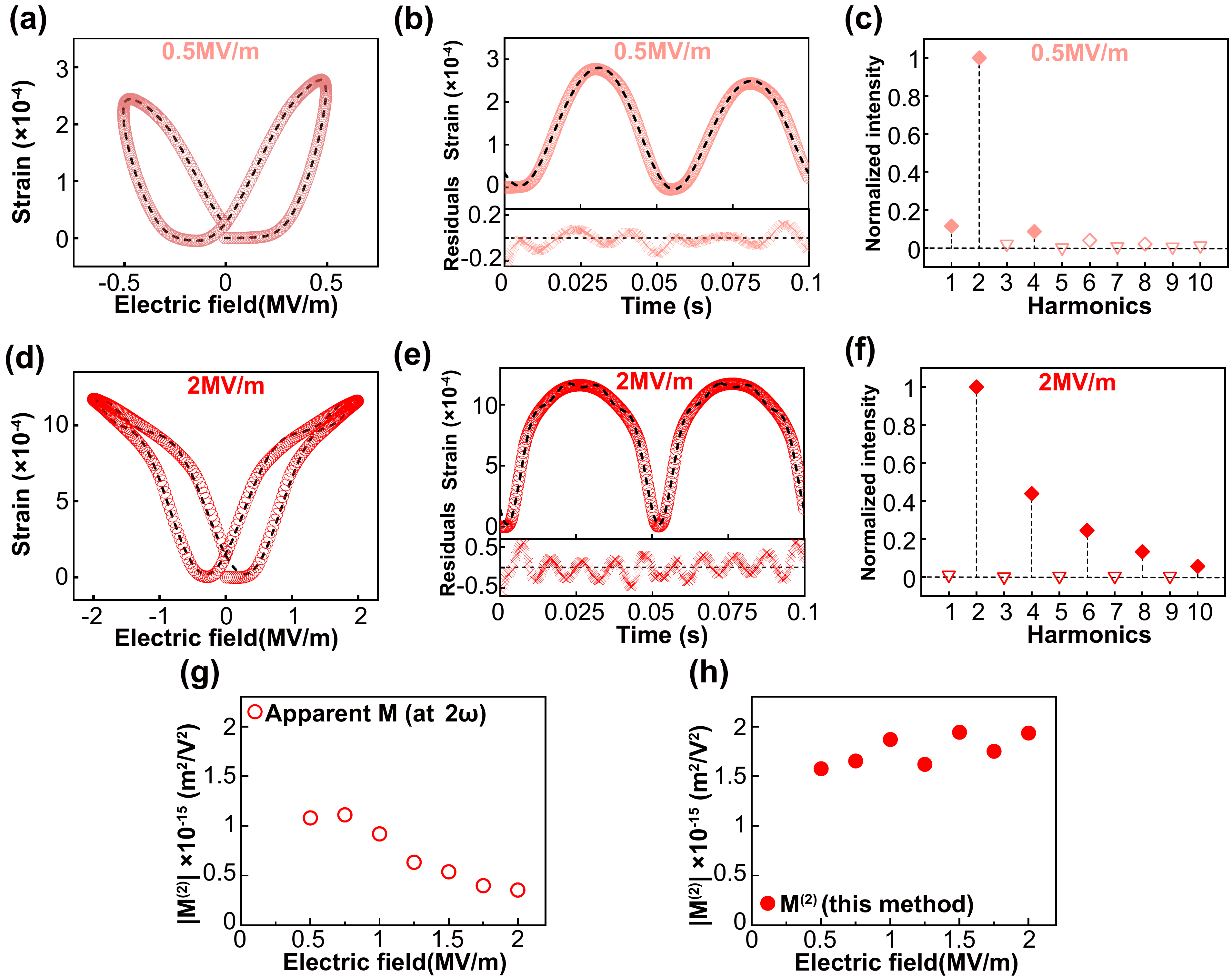}
    \caption{Nonlinear strain responses of PMN-10PT ceramic as a function of electric field (a) and time (b) in the low-field region (0.5\,MV/m) and high-field region (2\,MV/m) (d and e, respectively). Amplitude spectra from the Fourier transform represent the weight of the $n$-th harmonics for low-(c) and high-fields(f). The dashed lines in (b) and (e) are the fitted strains using Eq.\ref{eq:fit}. The hollow symbols in (c) and (f) represent the discarded harmonics, and the fit is carried out considering only the contribution from the harmonics represented by solid symbols. \JC{(g) The ``apparent'' $M$ coefficient, calculated from the deformation at 2$\omega$ (therefore including higher-order contributions) decreases as a function of the electric field whereas (h) the $M$ coefficient (corresponding to pure electrostriction) remains constant at about $1.76(\pm0.15)\times10^{-15}\,\rm m^2V^{-2}$ for all field amplitudes.}}
    \label{fig:PMNPT}
\end{figure*}

In order to test Li's description based on low-field and high-field regions, \textit{ac} electric fields between 0.5\,MV/m \PEJ{a value approximatively equal to the coercive field} (Figures \ref{fig:PMNPT}(a-c)) and 2\,MV/m (Figures \ref{fig:PMNPT}(d-f)) are applied to the PMN-10PT sample of Figures \ref{fig:Fig1}(a-c).
Figures \ref{fig:PMNPT}(a) and (d) show the strain that develops under the minimum and maximum amplitudes of electric fields considered, indicating that electrostriction is the predominant electromechanical coupling in PMN-10PT ceramics regardless of the field amplitude, as is expected.
Correspondingly, Figures \ref{fig:PMNPT}(b) and (e) show that the strain is positive and evolves, as a first approximation, at twice the frequency of the electric field (the strain is measured over one period of the electric field).
The normalized amplitude spectra calculated by Fourier analysis for each field amplitude are presented in Figures \ref{fig:PMNPT}(c) and (f). 
At low field, Figure \ref{fig:PMNPT}(c) indicates that the second harmonic predominates (as expected) but also that the first (piezoelectric) and fourth harmonic ($M^{(4)}$) contributions are not negligible.
The dashed line in Figure \ref{fig:PMNPT}(b) represents the result of a sinusoidal fit of $x(t)$ using Eq.\eqref{eq:fit} with linear, quadratic, and quartic terms ($n=1,2,4$) and is a satisfactory reproduction of the measured strain with little residuals.
%The red vectors of these non-negligible harmonics in Figure \ref{fig:PMNPT}(c) represent the phase angles of the strain \JC{with respect to the values} given by Eq.\eqref{eq:phi}. \PE{Here we need to expand to explain that more.}
%For example, the phase angle of the fourth harmonic $\varphi_4$ is the addition of the delay of the quartic electromechanical interaction $\psi_4$ and \PE{$\varphi_4=$}$\pi$, which indicates the negative sign of $M^{(4)}$ derived from Eq.\eqref{eq:x1(t)}.
%The electrostrictive coefficient $M$ is positive since the phase angle of the second harmonic ($\varphi_2$) is equal to the delay of the quadratic electromechanical interaction $\psi_2$. 
%It should be noted that $\varphi_2$ corresponds to the hysteresis of strain-field response.
%Similar to the loss due to the domain wall motion in the typical piezoelectric ceramic case\cite{hagimura1989impurity,uchino2006loss}, the value of $\varphi_2$ is related to the process of polarization reorientation of the different phase structures.

In the high-field region, the comparison of Figure \ref{fig:PMNPT}(c) and (f) shows that the normalized intensities of higher-order harmonics increase. This is an expected consequence of increasing the maximum field in a non-linear system.
%Figure \ref{fig:PMNPT}(c) shows the nonlinear strain when plotted as a function of time. 
Only keeping the second, fourth, sixth, eighth, and tenth harmonics in Figure \ref{fig:PMNPT}(f), this saturation curve can also be satisfactorily reproduced by fitting $x(t)$ Eq.\eqref{eq:fit} with $n$=2,4,6,8, and 10 \JC{(see Supplementary Information, Figure S1)}. %\PE{Here, we have oscillations of the residuals. This suggests than even higher harmonics should be considered. Does the 12th harmonic improve the fit?}
%As to the phase angles of these harmonics, we shall obtain the alternating positive-negative sequence of the signs \JC{shown in Figure\ref{fig:PMNPT}(f)}.

The above results have several consequences: (1) the calculated $M$ in the low-field region is \JC{$1.58(\pm0.04)\times10^{-15}\,\rm m^2V^{-2}$, consistent with the value calculated in the high-field region: $1.93(\pm0.03)\times10^{-15}\,\rm m^2V^{-2}$, when the higher-order terms are taken into account.} 
Hence, the $M$ remains constant (within less than \JC{20\%}) regardless of the amplitude of the driving field, \JC{as shown in Figure \ref{fig:PMNPT}(h)}. 
This is the expected behavior of a ``constant'' \JC{(see Supplementary Information about the evolution of $M^{(n)}$, Figure S2)}. 
(2) The low-field and high-field classification does not even ensure that PMN-10PT is purely electrostrictive (i.e., with no higher-order electromechanical terms), as even in the low-field region, the contribution of the quartic order interaction is not negligible. If only the strain at 2$\omega$ is considered (by filtering the Fourier spectrum to consider only this harmonic) then the $M$ coefficient value is 1.09$\times$10$^{-15}$\,m$^2$V$^{-2}$ at low field and 0.37$\times$10$^{-15}$\,m$^2$V$^{-2}$ at high field, \JC{as shown in Figure \ref{fig:PMNPT}(g)}. Such apparent field dependence of the $M$ coefficient is therefore due to the omission of higher-order coupling coefficients. %the method is fft and ifft get the 2w, then eliminate the phase differences, and polyfit x(t=2w) as a function of E}

Therefore, in PMN-10PT, the saturation of the polarization causes the saturation of the maximum electromechanical strain. Such a saturation should not be attributed to a field dependence of the $M$. In addition, calculating the $M$ coefficient from a filter centered on the $2\omega$ frequency leads to erroneous values. 

From a microscopic perspective, the active PNRs behave as ``seeds'' to induce the macro-domain from the dielectric matrix and then facilitate the switching of the macro-domains. 
The precise description of this dynamic process reflects the nonlinear electrostriction-based nature of relaxor ferroelectrics\cite{wang2018evolution}. 

\subsection{Harmonic analysis of non-classical electrostrictors}

On the contrary to ``classical electrostrictors'', some materials do not follow the empirical proxy based on elastic and dielectric properties\cite{newnham1997electrostriction}. 
When the deviation exceeds one order of magnitude, these materials have been defined as ``giant'' or ``non-classical'' electrostrictors\cite{yu2022defining}. 
Yavo \textit{et al.} have investigated the saturation of electrostriction in doped ceria ceramics, the first non-classical electrostrictor reported, pointing to charge trapping, voltage redistribution at the grain boundaries, and electrodes as origins of the observed saturation\cite{yavo2018relaxation,kabir2019effect,varenik2020dopant,varenik2023lead}. 
\ch{La2Mo2O9}-based materials (LAMOX) have also been reported to exhibit non-classical electrostriction \cite{li2018giant,yu2022}. Hereafter, we report results on ceramics of three of these LAMOX materials \JC{(structural and microstructural information can be found in the Supplementary Information, Figure S3 and S4)}. 

In contrast to the classical electrostrictor case, the maximum strain of substituted \ch{La2Mo2O9} ceramics, \textit{i.e.} for non-classical electrostrictors, exhibits a similar non-linear evolution when plotted as a function of $E^2$ and $P^2$ (see Figure \ref{fig:Fig1}(e) and (f)).
 %Figures \ref{fig:Fig1}(e) and (f) show the similar evolution of the maximum strain vs $E^2$ and vs $P^2$ for substituted \ch{La2Mo2O9} ceramics, \textit{i.e.} for non-classical electrostrictors. 
This is consistent with the linear dielectric character (Figure \ref{fig:Fig1}(d)) of LAMOX ceramics that allows to link the two electrostrictive coefficients to be made through $M=Q\varepsilon^2$ with $\varepsilon$ the permittivity.

\ch{La2Mo2O9} and W-substituted \ch{La2Mo2O9} ceramics were reported to exhibit non-classical electrostrictive effects that can be enhanced at elevated temperatures\cite{li2018giant}. 
It was further revealed that oxygen vacancy configurations and hopping dynamics cause the distortion of the local structure without measurable bulk polarization\cite{li2018giant}. 
Here, we shall extend the study of this family of materials by considering other substitutions.

Note that the curves of $x$-$E^2$ (Figure \ref{fig:Fig1}(e)) and $x$-$P^2$ (Figure \ref{fig:Fig1}(f)) of Gd-substituted (\ch{La_{1.8}Mo2Gd_{0.2}O9}) and (Gd,W)-substituted \ch{La2Mo2O9} (\ch{La_{1.7}Mo_{0.8}Gd_{0.3}W_{1.2}O9}) saturate, whereas the Nd-substituted \ch{La2Mo2O9} (\ch{La_{1.4}Mo2Nd_{0.6}O9}) curves shows the opposite tendency \textit{i.e.}, they tend to \emph{increase} under increasing electric fields.  
Clearly, the explanation posited for classical electrostrictors (that the saturation of the polarization causes the saturation of the strain under electric field) can not be applied to non-classical (``giant'') electrostrictors such as LAMOX as they remain linear dielectrics (see Figure \ref{fig:Fig1}(d)).
Even though these ``giant'' electrostrictors have been studied from different angles\cite{das2018anisotropic,asher2019computational,trujillo2022data}, only hypotheses have been put forward to explain their strain saturation. We show below that such behavior is explained when higher-order electromechanical coefficients are considered, thereby providing a unified description of the electromechanical strain saturation in both classical and non-classical electrostrictors.

\begin{figure}[htbp]
    \centering
    \includegraphics[width=0.48\textwidth]{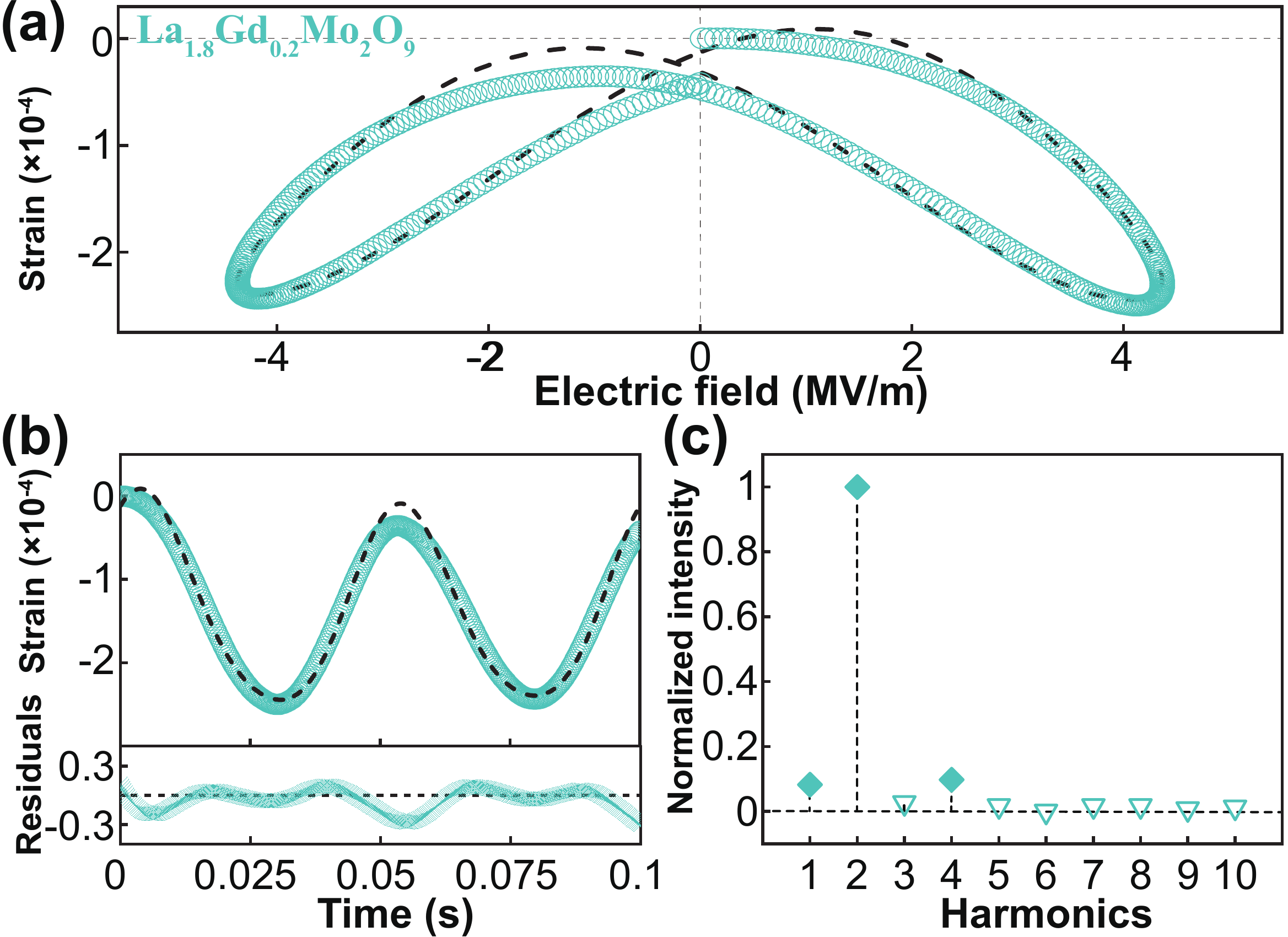}
    \caption{Negative electrostrictive response of Gd-substituted \ch{La2Mo2O9} ceramic as a function of the electric field (a) and time (b). The sinusoidal fit (dashed line in (b)) using Eq.\ref{eq:fit} reproduces the measured strain (blue line) by only considering the orders corresponding to harmonics which are identified from the normalized intensity of the fast Fourier transform of $x(t)$ (c). The opposite signs of $M=-1.89\times10^{-17}\,\rm m^2V^{-2}$ and $M^{(4)}$=$+3.27\times10^{-31}\,\rm m^4V^{-4}$ induce the saturation of the electrostrictive strain.}
    \label{fig:Gd}
\end{figure}

Figure \ref{fig:Gd}(a) shows the field-induced strain in Gd-substituted \ch{La2Mo2O9} indicative of a predominant negative electrostriction. % in Gd-substituted \ch{La2Mo2O9} ceramic (10\,Hz and 20$^\circ$C). 
Figure \ref{fig:Gd}(b) presents the time-dependence of the strain, measured over one field period and exhibiting at first glance an evolution at twice the field frequency. Both of these curves appear to point to a predominant electrostrictive behavior.
However, the amplitude spectrum (Figure \ref{fig:Gd}(c)) shows that, in addition to the (expected) predominant second harmonic, the first and fourth harmonics are not negligible.  
%The difference between the phase angle of the second harmonic $\varphi_2$ and the delay of the quadratic electromechanical interaction $\psi_2$ is $\pi$, which indicates the negative sign of the second harmonic. 
%The sign of the fourth harmonic is \JC{found to be} positive through the use of the same procedure.
The fitting of the $x(t)$ curve (represented by the dash line in Figure \ref{fig:Gd}(b)) provides a negative value for $M=-1.89(\pm0.02)\times10^{-17}\,\rm m^2V^{-2}$ and a positive value for $M^{(4)}$=$+3.27(\pm0.14)\times10^{-31}\,\rm m^4V^{-4}$. Therefore, as the field increases, the positive $M^{(4)}E_0^4$ term will compensate more and more for the negative $ME_0^2$ term. The saturation of the maximum strain in Gd-substituted \ch{La2Mo2O9} is therefore due to the opposite signs of $M$ and $M^{(4)}$. 

Another consequence of taking into account the fourth harmonic is that it leads to a value that differs from the one that would be obtained if only the response at $2\omega$ was considered ($-0.96\times10^{-17}\,\rm m^2V^{-2}$). This underlines, as was the case with classical electrostrictors, the importance of taking into account higher-order electromechanical terms to accurately evaluate the values of electrostrictive coefficients.
%After excluding the influence of $M^{(4)}$, the electrostrictive coefficient $M$ is $(-1.24\pm0.11)\times10^{-17}\,\rm m^2V^{-2}$ rather than $(-0.96)\times10^{-17}\,\rm m^2V^{-2}$ \JC{that one would obtain by carrying out a} simple quadratic fitting.
It is worth pointing out that the non-negligible first harmonic does not necessarily correspond to piezoelectricity. Indeed, the last measured strain point differs from the first measured point (see Figure \ref{fig:Gd}(a), which can also be seen in the residuals in Figure \ref{fig:Gd}(b)). 
\JC{This can be due to various phenomena, e.g., to the damping behavior of defect dipoles or to the initial alignment of dipoles during the first application of the electric field.}
This leads to a non-zero overall slope of the strain as a function of time that contributes to the non-zero first harmonic. %\PE{Wouldn't it also "pollute" the higher-order harmonics?} \JC{Yes, but there is no obvious contribution of higher harmonics in the plot.}
%The second piece of evidence is that $\varphi_1$ is negative (counterclockwise), which contradicts the delaying response of strain with regard to the field.

\begin{figure}[htbp]
    \centering
    \includegraphics[width=0.48\textwidth]{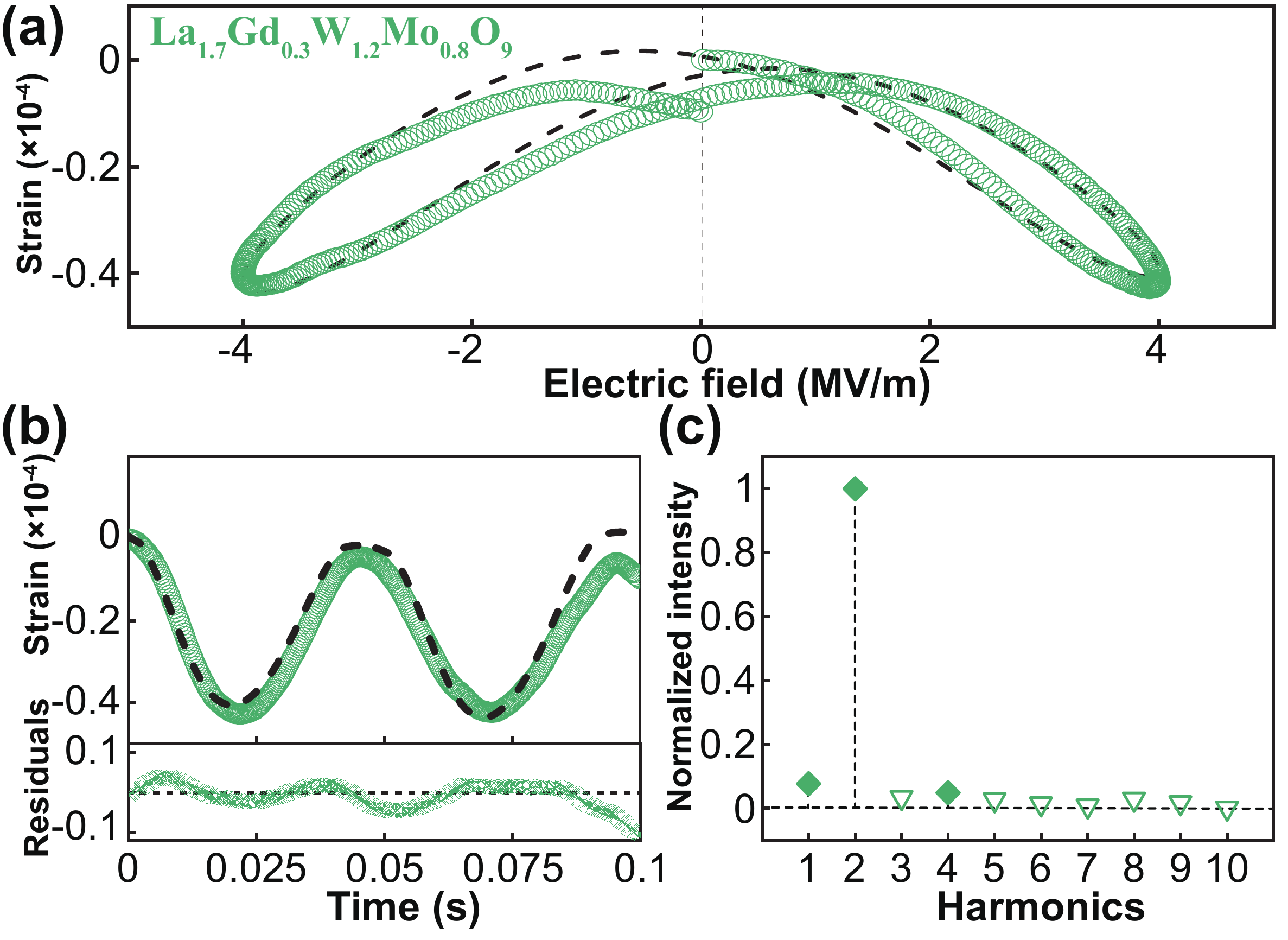}
    \caption{Negative electrostrictive response of (Gd-W)-substituted \ch{La2Mo2O9} ceramic as a function of the electric field (a) and time (b). The sinusoidal fit (dashed line in (b)) using Eq.\ref{eq:fit} reproduces the measured strain (blue line) by only considering the orders corresponding to harmonics which are identified from the normalized intensity of the fast Fourier transform of $x(t)$ (c). The opposite signs of $M=-0.35\times10^{-17}\,\rm m^2V^{-2}$ and $M^{(4)}=+5.53\times10^{-32}\,\rm m^4V^{-4}$ induce the saturation of the electrostrictive strain.}
    \label{fig:GdW}
\end{figure}

If the additional addition of tungsten in Gd-substituted \ch{La2Mo2O9} enables to reduce the electrostrictive losses (the strain vs electric field curve is less hysteretic than without tungsten, compare Figure \ref{fig:Gd}(a) and Figure \ref{fig:GdW}(a)), it comes at the cost of decreasing significantly (5-fold reduction) the maximum strain. The $M$ coefficient of (Gd,W)-substituted \ch{La2Mo2O9} ($M=-0.35(\pm0.01)\times10^{-17}\,\rm m^2V^{-2}$), considering the first, second, and fourth harmonic is therefore lower than the one of Gd-substituted \ch{La2Mo2O9}. 
%There are several results of adding extra tungsten in Gd-substituted \ch{La2Mo2O9}: diminishing the amplitude of strain response ($M=(-0.31\pm0.05)\times10^{-17}\,\rm m^2V^{-2}$) as shown in Figure \ref{fig:GdW}(a), reducing effectively the electrostrictive loss ($\varphi_2$) shown as in Figure \ref{fig:GdW}(c).
As in the case of Gd-substituted \ch{La2Mo2O9}, the apparent saturation of electrostriction remains due to the opposite signs of $M$ and  $M^{(4)}$ ($+5.53(\pm0.56)\times10^{-32}\,\rm m^4V^{-4}$).

\begin{figure}[htbp]
    \centering
    \includegraphics[width=0.48\textwidth]{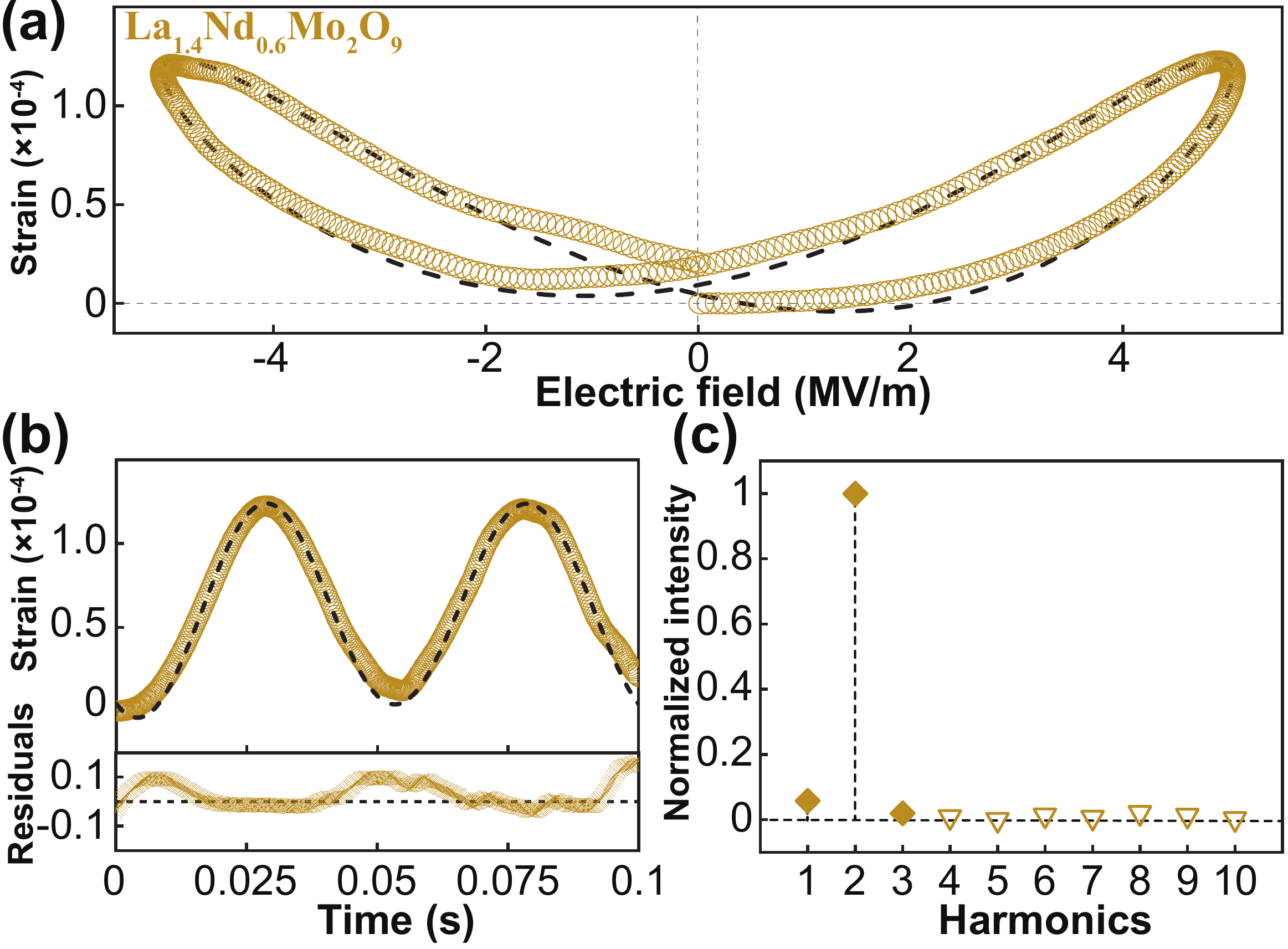}
    \caption{Positive electrostrictive response of Nd-substituted \ch{La2Mo2O9} ceramic as a function of the electric field (a) and time (b). The sinusoidal fit (dashed line in (b)) using Eq.\ref{eq:fit} reproduces the measured strain (blue line) by only considering the orders corresponding to harmonics which are identified from the normalized intensity of the fast Fourier transform of $x(t)$ (c). The similar signs of $M=+0.49\times10^{-17}\,\rm m^2V^{-2}$ and $M^{(3)}=+5.36(\pm1.27)\times10^{-26}\,\rm m^3V^{-3}$ induce the increase of the electrostrictive strain when the field amplitude increases.}
    \label{fig:Nd}
\end{figure}

We now turn to Nd-substituted \ch{La2Mo2O9}. This material exhibits a positive electrostrictive strain (see Figure \ref{fig:Nd}(a)) with an electrostrictive coefficient $M=+0.49(\pm0.01)\times10^{-17}\,\rm m^2V^{-2}$ similar in amplitude but opposite in sign to the one of (Gd,W)-substituted \ch{La2Mo2O9}. 
In addition to the second harmonic, the first, and third harmonics are considered from the amplitude spectrum (Figure \ref{fig:Nd}(c)) in the fitting of $x(t)$ (Figure \ref{fig:Nd}(b)) using Eq.\ref{eq:fit} when $n=1,2,3$. 
%The phase diagrams of second and fourth harmonics in Figure \ref{fig:Nd}(c) exhibit both positive signs, giving rise to the \JC{apparent} divergent behavior \JC{of the electrostrictive coefficients shown} in Figure \ref{fig:Fig1}(c) and (d).
In contrast to the previous cases, the quartic order term is negligible, while the $M^{(3)}$ coefficient ($+5.36(\pm1.27)\times10^{-26}\,\rm m^3V^{-3}$) exhibits a positive sign. As a consequence, rather than compensating each other and causing the electromechanical strain to saturate as the field increases, the electromechanical strain of Nd-substituted \ch{La2Mo2O9} \emph{increases} as the field amplitude increases. The modest relative weight of the harmonics considered compared to the amplitude of the second harmonic explains the modest positive deviation from linearity over the investigated field range. In addition, as $M^{(3)}$ does not contribute to the signal at 2$\omega$ (cf. Eq.\ref{eq:x1(t)}), the value of $M$ one would get considering only the second harmonic (0.44$\times$10$^{-17}$m$^2$V$^{-2}$) is close (within 10\%) to the actual one. However, not considering this higher-order coupling prevents from explaining the increase of the electromechanical strain when the field increases. % \PE{Can you, here as well, provide the value one would get of the $M$ coefficient if only the second harmonic was considered?}\JC{0.44e-17}

The conclusion to draw from these substituted \ch{La2Mo2O9} is that substitution elements can tune the amplitude, hysteresis, and saturation behavior of the electromechanical properties in non-classical electrostrictors, through the contributions of higher-order electromechanical coupling coefficients. Considering higher-order harmonics based on the amplitude spectra of their Fourier transform enables in non-classical electrostrictors as in classical electrostrictors to calculate a field-independent and more accurate values of the electrostrictive coefficients. We anticipate that microstructural analyses of these substituted \ch{La2Mo2O9} will enable to correlate the signs of the higher-order coupling coefficients to microstructural features.

\section{Conclusion}

The study of classical and non-classical electrostrictive materials requires the consideration of higher-order electromechanical terms, even at low fields, to calculate the actual values of electrostrictive coefficients.
The apparent field-dependence of electrostrictive coefficients calculated either from a quadratic fit of the strain versus electric field curves or after having filtered only the second harmonic may explain the dispersion of electrostrictive coefficients reported in the literature. 
Indeed, such a procedure does not enable the determination of an accurate value of the electrostrictive coefficients.
In addition, the signs of the higher-order coefficients enable predictions to be made about the behavior of the maximum strain under increasing electric field. 
The saturation (or increase) of the electromechanical strain under electric field is explained by the increasing contribution of higher-order couplings in the electromechanical response.

We propose a scheme to determine the actual values of the electrostrictive coefficients and predict the evolution of electromechanical strain under electric field: first, the amplitude spectrum of the time-dependent strain curve is calculated from its Fourier transform, then the harmonics to consider are chosen based on the histogram of this amplitude spectrum, and finally are used to carry out a fit of the time evolution of the strain with a polynomial function of $\sin(\omega t)$. This provides values of electrostrictive coefficients that do not depend on the amplitude of the applied electric field. In addition, from the sign of the higher-order electromechanical coefficients, it is possible to determine whether they will compensate each other (if they have opposite signs) or add to each other (if they have the same sign) and therefore predict whether the strain will saturate or increase when the field increases.

% We propose an approach to calculate the electrostriction \JC{(or piezoelectric)} coefficients, which can be strongly underestimated or overestimated by the presence of higher-order harmonics. 
% \JC{For} electrostriction, the precise calculation of the $M$ coefficient can be used as a benchmark to compare electrostrictive strains generated in \JC{dielectrics}.
% \JC{The} $M$ coefficients remain constant, as shown in the example of PMN-10PT under the different amplitudes of the fields\JC{, rather than vary depending on the field amplitude}.
% More importantly, we thoroughly investigate the \JC{apparent} saturation or anti-saturation of the \JC{electrostrictive coefficient values} caused by the counteracting or contributing higher-order harmonics from a simple \textit{ac} field-driven measurement. 
% The response of Nd-substituted \ch{La2Mo2O9} confirms that it is possible to reduce or even reverse the tendency of saturation by
% using chemical doping routes.  
% Our finding thus suggests that the analysis of the electromechanical nonlinearities provides \JC{more reliable values and potentially novel insight about the electromechanical coupling} that are often neglected by \JC{a simpler approach}.
 
\medskip
\textbf{Acknowledgements}
This work was partially financed by ANR-19-ASTR-0024-01 and ANR-20-CE08-0012-1 grants 

\medskip

% Use the following code if you wish to generate your bibliography with BibTeX;
% replace the string "MSP-template" below with the name(s) of
% the BibTeX data base(s) you want to use.
% The resulting bibliography-output (the content of the .bbl file)
% must be pasted back into this file before submission.
% Please also include your BibTeX data base file(s) in your submission
% so that we can re-run BibTeX if necessary.
%
\bibliographystyle{MSP}
\bibliography{ref}

\end{document}